\documentclass[11pt,twoside]{article} 
\usepackage{asp2004}
\usepackage{epsf}
\usepackage{psfig}
\usepackage{lscape} 
\usepackage{amsmath,amstext,amsfonts,amssymb}

\begin{document} 
\title{FUSE spectroscopy of PG~1159 Stars}
\author{E. Reiff,$^1$ T. Rauch,$^{1,2}$ K. Werner,$^1$ and J. W. Kruk$^3$}
\affil{$^1$Institut f{\"u}r Astronomie und Astrophysik, Universit{\"a}t
  T{\"u}bingen, Sand\,1, 72076 T{\"u}bingen, Germany \\
$^2$Dr.-Remeis-Sternwarte, Universit{\"a}t Erlangen-N{\"u}rnberg,
Sternwart\-strasse 7, 96049 Bamberg, Germany \\  
$^3$Department of Physics and Astronomy, The Johns Hopkins University,
Baltimore, MD 21218, USA }

\begin{abstract} 
PG~1159 stars are hot hydrogen-deficient post-AGB stars with effective
temperatures within a range from 75\,000\,K up to 200\,000\,K. These stars are
probably the result of a late helium-shell flash that had occurred during their
first descent from the AGB. The lack of hydrogen is caused by flash-induced
envelope mixing and burning of H in deeper regions.  Now the former intershell
matter is seen on the surface of the stars. Hence the stellar atmospheres show
metal abundances drastically different from the solar values.  Our sample
comprises ten PG~1159 stars with effective temperatures between 85\,000\,K and
170\,000\,K. We present first results of our spectral analysis based on FUV
spectra obtained with the Far Ultraviolet Spectroscopic Explorer (FUSE).
\end{abstract}

\begin{figure}[!ht]
\plotone{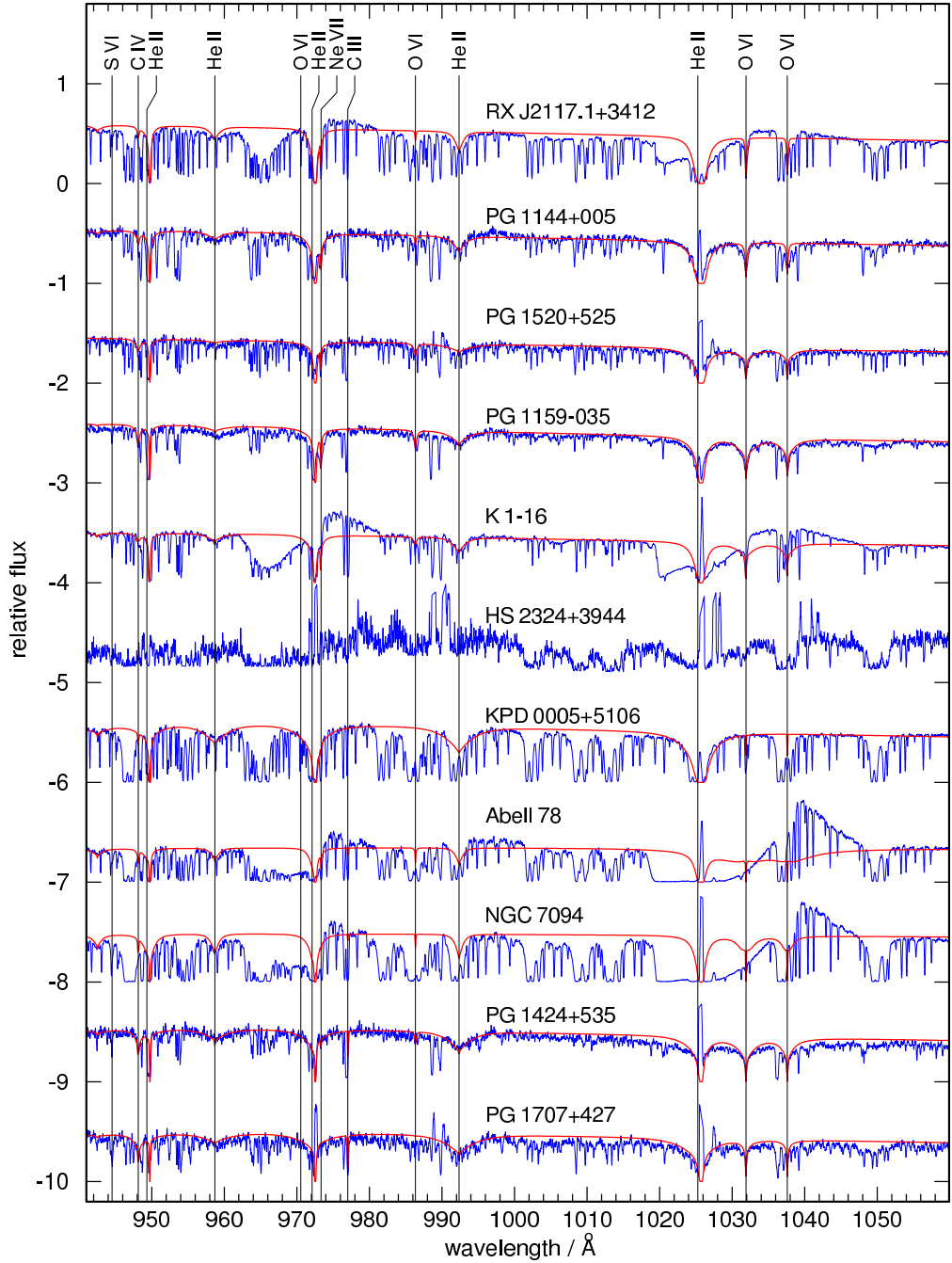}
\caption{Detail of FUSE spectra compared with
  models, except for HS\,2324+3944.  Some
  photospheric line identifications are given. Note the newly identified
  Ne\,{\scriptsize VII}~973\,\AA\ line. Some stars show P~Cyg
  profiles of the C\,{\scriptsize III} and O\,{\scriptsize VI} resonance
  lines, which are not fit by our static models. All other
  absorption features which are not reproduced by our models are interstellar.}
\end{figure}

\section{Introduction}
We are analysing FUSE spectra from ten PG~1159 stars and, in addition, the
hottest known DO white dwarf KPD~0005+5106, which we have obtained through own
PI observations or from the MAST archive. They allow an independent
investigation of the atmospheric parameters, which were previously determined
from optical and/or HST UV spectroscopy, in some cases with low
precision. Table~1 summarizes the currently known parameters, as taken from
literature, which we use as a starting point for our analysis. Our
plane-parallel static models used here are calculated with TMAP, the T\"ubingen
NLTE Model Atmosphere Package (Werner et al\@. 2003, Rauch \& Deetjen 2003).

\section{Observations and First Results}
In Fig.\,1 we display details of the FUSE spectra of our sample ordered by
effective stellar temperature, beginning with  the hottest object on the
top. The FUV spectra of PG~1159 stars are dominated by lines from
He\,{\scriptsize II}, C\,{\scriptsize IV}, and O\,{\scriptsize VI}. In addition,
several objects show S\,{\scriptsize VI} lines at 933.38\,\AA\ and
944.52\,\AA. Miksa et al.\ (2002) detected sulfur in the spectra of PG\,1159
stars for the first time by identifying this resonance doublet in the
FUSE spectrum of K\,1-16. One of the strongest absorption lines in the FUSE
spectra, located at 973.3\,\AA, remained unidentified until we found recently
that it stems from Ne\,{\scriptsize VII} (Werner et al. 2004a). This enables to
determine the neon abundance, which is otherwise difficult or impossible to do
(Werner et al. 2004b). Another new identification is the detection of highly
ionised fluorine lines (Werner et al.\ 2005). During the
course of our current analysis we identified in addition for the first time
lines from silicon and phosphorus (Si\,{\scriptsize IV},  P\,{\scriptsize
V}), see Fig.\,2. Abundance analyses from these features are being
performed.
\begin{figure}[t]
\plotone{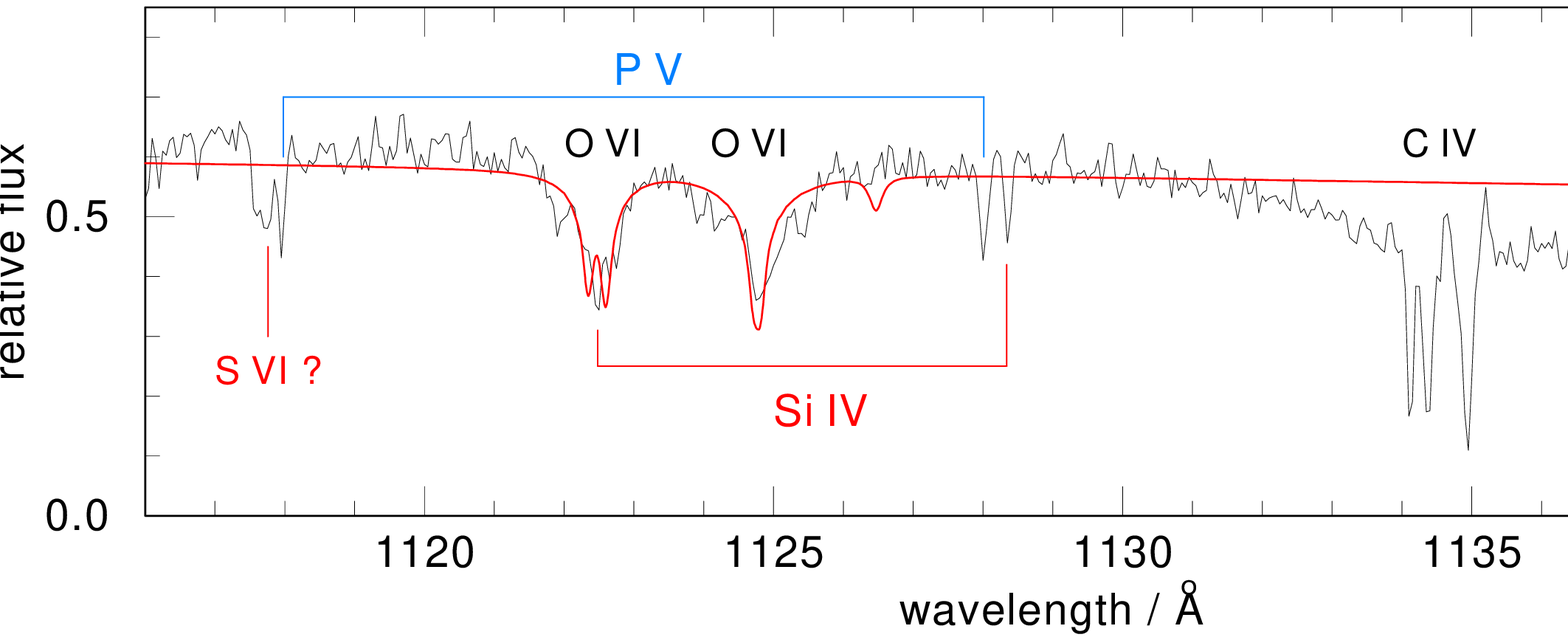}
\caption{First identification of phosphorus and silicon in a PG~1159 star
  (PG\,1424+535). Note also the presence of a strong fluorine line 
(Werner et al\@. 2005), and possibly a sulfur line (the
S\,{\scriptsize VI} resonance doublet is clearly detected). The P, Si,
and S lines are not included in the model, as well as the broad
C\,{\scriptsize IV} trough.} 
\end{figure}

\begin{table}[!h] 
\caption{Atmospheric parameters of our objects, taken
  from the literature. $E_{\mathrm{B-V}}$ was derived from our FUSE spectra. 
 All objects are PG~1159 stars, except the DO white dwarf KPD 0005+5106.}
\begin{center}
\begin{tabular}{l r c c c r r r l l}
\tableline
\noalign{\smallskip}
Object & $T_{\mathrm{eff}}$ & $\log g$ & H & He
&\multicolumn{1}{c}{C} &\multicolumn{1}{c}{O} & Ne &
$E_{\mathrm{B-V}}$ & Ref.\\
\noalign{\smallskip}
\cline{4-8}
\noalign{\smallskip}
      & $[$kK$]$     & [cgs]  &\multicolumn{5}{c}{[mass fractions]}& &\\ 
\noalign{\smallskip}
\tableline
\noalign{\smallskip}
RXJ2117+3412     & 170 & 6.0 &      & 38.0 & 56.0 &  6.0 & 2.0 & 0.03 & 1,2\\ 
PG\,1144+005     & 150 & 6.5 &      & 39.0 & 58.0 &  1.6 &     & 0.01 & 3\\    
PG\,1520+525    & 150 & 7.5 &      & 44.0 & 39.0 & 17.0 &     & 0.00 & 4\\     
PG\,1159-035     & 140 & 7.0 &      & 33.0 & 49.0 & 17.0 &     & 0.00 & 4,5\\ 
K 1-16           & 140 & 6.4 &      & 33.0 & 50.0 & 17.0 & 2.0 & 0.025& 6,2\\  
HS\,2324+3944    & 130 & 6.2 & 21.  & 41.0 & 37.0 &  1.0 &     &      & 7\\ 
KPD\,0005+5106   & 120 & 7.0 &      & 99.4 &  0.4 &      &     & 0.025& 1\\    
Abell\,78        & 110 & 5.5 &      & 33.0 & 50.0 & 15.0 &     & 0.13 & 8,9 \\ 
NGC\,7094        & 110 & 5.7 & 36.  & 43.0 & 21.0 &      &     & 0.11 & 10\\   
PG\,1424+535     & 110 & 7.0 &      & 50.0 & 44.0 &  6.0 &     & 0.02 & 4\\ 
PG\,1707+427     &  85 & 7.5 &      & 43.0 & 38.5 & 17.0 &     & 0.00 & 4\\
\noalign{\smallskip}
\tableline  
\end{tabular}
\end{center}
\vspace*{-8pt}
\begin{small}
\begin{tabular}{l l}
(1) Werner et al. (1996)             & (6) Kruk \& Werner (1998) \\
(2) Werner \& Rauch (1994)           & (7) Dreizler et al. (1996) \\
(3) Werner \& Heber (1991)           & (8) Werner \& Koesterke (1992) \\
(4) Dreizler \& Heber (1998)         & (9) Werner et al. (2003) \\
(5) Werner et al. (1991)  & (10) Dreizler et al. (1997)
\end{tabular}
\end{small}
\end{table}

So far, our synthetic spectra reproduce the FUSE spectra already quite
well, as can be seen in Fig.\,1. This allows us to derive the
interstellar reddening from the continuum shape. The results are
listed in Table~1. The next step in our analysis will be to
investigate the abundance dependences of the metal lines lines in
order to fine-tune the atmospheric parameters.

Some of our analysed objects, namely RX\,J2117.1+3412, K\,1-16,
Abell\,78 and NGC\,7094, show P\,Cygni profiles of the C\,{\scriptsize
III} and O\,{\scriptsize VI} resonance lines (see Fig.\,1). For three
of these stars (RX\,J2117.1+3412, K\,1-16 and NGC\,7094) mass-loss
rates and terminal velocities were determined in a former study by
Koesterke et al.\ (1998) based on HST spectra. Abell\,78 has been
investigated as well in a recent analysis of FUSE and ORFEUS spectra
by Herald \& Bianchi (2004). Mass-loss from these stars will be
accounted for in our future analyses with expanding atmosphere models.

\acknowledgements{This work is supported by the DLR under grant \linebreak[4]
  50\,OR\,0201 (T.R.), and the FUSE project, funded by NASA contract
  NAS5-32985 (J.W.K.).}


\begin{references}

\reference Dreizler, S., \& Heber, U. 1998, \aap, 334, 618

\reference Dreizler, S., Werner, K., Heber, U., \& Engels, D. 1996,
\aap, 309, 820

\reference Dreizler, S., Werner, K., \& Heber, U. 1997, in IAU Symp. 
180, Planetary Nebulae, eds. H. J. Habing \& H. J. G. L. M. Lamers
(Dordrecht: Kluwer), 103

\reference Koesterke, L., Dreizler, S., \& Rauch, T. 1998, \aap, 330, 1041

\reference Kruk, J. W., \& Werner, K. 1998, \apj, 502, 858 

\reference Herald, J. E., \& Bianchi, L. 2004, \apj, 609, 378

\reference Miksa, S., Deetjen, J.L., Dreizler, S., et al. 2002, \aap, 389, 953

\reference Rauch, T., \& Deetjen, J. L\@. 2003,
           in: Stellar Atmosphere Modeling,
           eds\@. I\@. Hubeny, D\@. Mihalas, K\@. Werner,
           ASP Conference Series, 288, 103 

\reference Werner, K., \& Heber, U. 1991, \aap, 247, 476	

\reference Werner, K., \& Koesterke, L. 1992, Lecture Notes in Physics,
Vol. 401, eds. U. Heber \& C.S. Jeffery (Springer: Berlin), 288

\reference Werner, K., \& Rauch, T. 1994, \aap, 284, L5

\reference Werner, K., Heber, U., \& Hunger, K. 1991, \aap, 244, 437

\reference Werner, K., Dreizler, S., Heber, U., et al. 1996, \aap, 307, 860

\reference Werner, K., Dreizler, S., Koesterke, L., \& Kruk, J.W.
2003, in IAU Symp. 209, Planetary Nebulae: Their Evolution and Role in
the Universe, eds. S. Kwok, M. Dopita, \& R. Sutherland (Astronomical
Society of the Pacific), 239

\reference  Werner, K., Dreizler, S., Deetjen, J. L., et al. 2003
           in: Stellar Atmosphere Modeling,
           eds\@. I\@. Hubeny, D\@. Mihalas, K\@. Werner,
           ASP Conference Series, 288, 31 

\reference Werner, K., Rauch, T., Barstow, M. A., \& Kruk,
J. W. 2004a, \aap, 421, 1169

\reference Werner, K., Rauch, T., Reiff, E., Kruk, J. W., \&
Napiwotzki, R. 2004b, \aap, in press

\reference Werner, K., Rauch, T., \& Kruk, J. W. 2005, \aap, submitted

\end{references}
\end{document}